\documentclass[10pt,journal,onecolumn]{IEEEtran}
\usepackage{graphicx}
\usepackage{dcolumn}
\usepackage{bm}

\usepackage{amsmath}
\usepackage{multirow}
\renewcommand{\arraystretch}{1.7}
\usepackage{amssymb}
\usepackage{url}
\usepackage{subfigure}
\usepackage{algorithm}
\usepackage{algorithmic}
\usepackage{enumerate}
\usepackage{slashbox}
\usepackage{float}
\usepackage{soul}
\usepackage[normalem]{ulem}
\usepackage{makecell}
\usepackage{slashbox,pict2e}

\newtheorem{theorem}{Theorem}

\usepackage{amsmath}
\ifCLASSOPTIONcompsoc
  \usepackage[nocompress]{cite}
\else
  \usepackage{cite}
\fi
\ifCLASSINFOpdf
\else
\fi

\begin{document}
%
\title{Evaluation of Community Detection Methods}
%
%
%
%

\author{Xin Liu, Hui-Min Cheng, Zhong-Yuan Zhang
\IEEEcompsocitemizethanks{\IEEEcompsocthanksitem Xin Liu, Hui-Min Cheng and Zhong-Yuan Zhang are with School of Statistics and Mathematics, Central University of Finance and Economics, P.R.China. \protect\\
E-mail: zhyuanzh@gmail.com
}
\thanks{Manuscript received April 19, 2005; revised August 26, 2015.}}

%
%

\markboth{Journal }%
{Shell \MakeLowercase{\textit{et al.}}: Bare Demo of IEEEtran.cls for Computer Society Journals}
%



\IEEEtitleabstractindextext{%
\begin{abstract}
Community structures are critical towards understanding not only the network topology but also how the network functions. However, how to evaluate the quality of detected community structures is still challenging and remains unsolved. The most widely used metric, normalized mutual information (NMI), was proved to have finite size effect, and its improved form relative normalized mutual information (rNMI) has reverse finite size effect. Corrected normalized mutual information (cNMI) was thus proposed and has neither finite size effect nor reverse finite size effect. However, in this paper we show that cNMI violates the so-called proportionality assumption. In addition, NMI-type metrics have the problem of ignoring importance of small communities. Finally, they cannot be used to evaluate a single community of interest. In this paper, we map the computed community labels to the ground-truth ones through integer linear programming, then use kappa index and F-score to evaluate the detected community structures. Experimental results demonstrate the advantages of our method.
\end{abstract}

}

\maketitle

\IEEEdisplaynontitleabstractindextext

%
\IEEEpeerreviewmaketitle

\section{\label{Introduction}Introduction}
There are often community structures in complex networks, which often refer to groups of nodes that are intra-connected tightly and inter-connected loosely. Community structures are very important towards understanding not only the network topology but also how the network functions, and many methods have been proposed to  detect community structures from different perspectives\cite{fortunato2010community}. However, how to evaluate them is still a challenging problem and remains unsolved.

The most widely used approach is testing and comparing different methods on synthetic benchmark networks with known community labels of nodes. Their performance is determined by calculating the similarity between computed labels and the ground-truth ones. The metrics are borrowed heavily from the field of clustering, including purity\cite{Manning2008Introduction}, rand index\cite{Rand1971Objective}, etc., and the most widely used one among them is normalized mutual information (NMI)\cite{Danon2005Comparing,Strehl2003Cluster}. However, recent research showed that NMI has finite size effect due to finite size of network\cite{zhang2015evaluating}. Hence relative normalized mutual information (rNMI)\cite{zhang2015evaluating} was proposed to handle this issue, but it has reverse finite size effect due to the same reason. To overcome the biases of NMI and rNMI, corrected normalized mutual information (cNMI)\cite{Lai2016A} was proposed. Actually, cNMI is not good enough either because it violates the proportionality assumption, as will be discussed later.

The main challenge facing designing reasonable indices for evaluation of community structures detection methods is the limited amount of information available from the community labels. The problem is kind of unsupervised learning, and the results can only provide the information that which nodes are clustered together and which ones are not, making no sense to point-wise compare between computed labels and ground-truth ones. 
For example, the two community label vectors $[1, 1, 1, 1, 1, 1, 2, 2, 3, 3]$ and $[2, 2, 2, 2, 2, 2, 3, 3, 1, 1]$ are actually identical. Furthermore, the indices cannot be used to evaluate the ability to detect a single community of interest due to the same reason, which is often very important in real applications such as understanding and assessing terror organizations, mining biological networks for unknown pathways, etc.
The above problems can be largely mitigated in the field of supervised learning since the computed labels are relatively more informative, making point-wise label comparison reasonable.
In this paper, we map the computed community labels to the ground-truth ones under the framework of linear programming, and then use kappa index and F-score for classification to evaluate community structures. Experimental results demonstrate the advantages of our method.

The rest of the paper is organized as follows: Sect. \ref{related_work} briefly reviews the representative indices for performance evaluation in clustering and classification; Sect. \ref{model} proposes the linear programming method, mapping the computed community labels to the ground-truth ones; Sect. \ref{experiments} demonstrates the advantages of our method over others empirically, and Sect. \ref{conclusion} concludes.

\section{\label{related_work}Related Work}
In this section, we will briefly review the representative indices for performance evaluation in clustering and classification.
\subsection{\label{clustering}Indices of clustering validity in unsupervised learning}
There are two types of indices for clustering evaluation: internal ones and external ones, and in this subsection we offer a brief analysis of some representative ones.
\subsubsection{\label{internal}Internal Indices}
Internal indices evaluate the clustering result without any external information and are only based on the information intrinsic
to the data, such as sum of squares, Calinski-Harabasz index and Silhouette index, etc.
\begin{enumerate}[i)]
\item Sum of squares:  As the simplest and most intuitive evaluation index, it computes the sum of within-cluster sum of squares as follows:
    \begin{equation}\label{eq:01}
    \sum^{c}_{r=1}\sum^{n}_{i=1}w_{ir}\left\|s_{i}-\bar{s}^{(r)}\right\|^2,
    \end{equation}
    where there are $n$ samples $s = \{s_{1},\, s_{2},\, \cdots,\, s_{n}\}$ with $c$ clusters, and $\bar{s}^{(r)},\, r=1,2,\dots,c$ is the cluster center, $w_{ir}$ is the clustering membership indicator, i.e., $w_{ir}=\begin{cases}\vspace{1mm}1,\quad s_{i}\in \mbox{cluster }  r\\0,\quad \mbox{otherwise}.\end{cases}$
\item Calinski-Harabasz index\cite{T1974A}: This index considers not only internal compactness of clusters, but also separability of clusters. It is very similar with $F$-test in ANOVA, and is calculated as the  following ratio of between-cluster mean and sum of within-cluster sum of squares:\\
\begin{equation}
\frac{BGSS/(c-1)}{WGSS/(n-c)},
\end{equation}
where \begin{equation*}
BGSS = \displaystyle\sum_{r=1}^cn_r\|\bar{s}^{(r)}-\bar{s}\|^2
\end{equation*}
and $WGSS$ is defined in Eq.(\ref{eq:01}), $n, c$ are the numbers of samples and clusters, respectively, $c-1$ and $n-c$ are degrees of freedom, respectively, $n_r$ is the number of samples in cluster $r$, and $\bar{s}$ is the center of the $n$ samples.

The larger the index, the better the clustering result is.
\item Silhouette index \cite{Rousseeuw1987Silhouettes}: The index for sample $i$ calculates how similar the sample is to its own cluster compared with other ones, and is defined as follows:\\
\begin{equation}
S\left(i\right)=\dfrac{b\left(i\right)-a\left(i\right)}
{\max\{a\left(i\right),b\left(i\right)\}},
\end{equation}
where $a\left(i\right)$ is the average distance of sample $i$ to the ones in its own cluster $r$, $b\left(i\right)= \min\{b\left(ik\right), \, k=1,\, 2,\, \cdots\, c,\, k\neq r\}$, and $b\left(ik\right)$ is the average distance of sample $i$ to those in the $k$th cluster ($k\neq r$).

The Silhouette index of clustering result is the average of $S(i), \, i=1,2,\cdots,n$, and ranges from -1 to +1.


 Since almost all internal indices are actually measuring the compactness of clusters, they naturally prefer the clustering result with larger cluster number and fewer samples in each cluster. For example, Calinski-Harabasz index, which is considered as one of the most effective indices\cite{kozak2012dendrite}, chooses 5 as the optimal cluster number on the famous Iris dataset, whose structures are simple and has 3 clusters \cite{Mufti2008Determining}.

 Furthermore, it is hard to extend the above indices to network data since there are no coordinates for the nodes, making it difficult to calculate the distance between nodes. Specially designed indices include modularity, partition density, etc.

\item Modularity \cite{newman2006modularity}: The basic idea of modularity is very similar to hypothesis testing: given a network partition, it compares the fraction of links inside each module of the partition with the so-called null model, i.e., the expected fraction of links inside the corresponding module in degree preserving counterpart, and sums the differences over all of the modules of the partition. Higher modularity value indicates more reasonable and meaningful partition. For a complex network $G=(V,E)$, if it is divided into $c$ communities, the modularity is defined as follows:\\
    \begin{equation}
       Q=\dfrac{1}{2m}\sum^{n}_{i,j=1}\left(A_{ij}-\dfrac{k_{i}k_{j}}
        {2m}\right)\delta\left(g_{i},g_{j}\right),
    \end{equation}
    where $m$ is the number of edges in the network, $g_{i}$ is the community label of node $i$, $g_{i}\in \{1,2,\dots,c\}$, $k_{i}$ is degree of node $i$. $A = (A_{ij})_{n\times n}$ is the adjacency matrix.

    Modularity can not only be used for evaluation of community structures, but also be optimized directly. However, modularity-optimization strategy has a resolution limit \cite{lancichinetti2011limits}. Modularity maximization usually tends to merge small communities into large ones in network.  Further studies have found more disadvantages of modularity \cite{Good2010Performance,Bagrow2012Communities,Zhang2009Modularity}. These drawbacks may lead to wrong estimation of the number of communities \cite{Kehagias2013Bad}. New community quality metrics, such as modularity density, were proposed to overcome these problems \cite{chen2015new}.
\item Partition density \cite{Ahn2010Link}: It only considers the compactness of communities, i.e., local information, and is defined as follows:\\
\begin{equation}
D = \dfrac{2}{m}\sum^{c}_{\alpha=1}m_{\alpha}\dfrac{m_{\alpha}-
\left(n_{\alpha}-1\right)}{\left(n_{\alpha}-2\right)\left(n_{\alpha}-1\right)},
\end{equation}
where $m_{\alpha}$ and $n_{\alpha}$ are the numbers of edges and vertices in the community $\alpha$, respectively.

Although partition density is free from the resolution limit of modularity, it suffers from the problem of inverse resolution limit and has preference towards triangles\cite{Lee2017Inverse}.
\end{enumerate}
In addition, Yang, et.al \cite{yang2015defining} studied the internal indices in depth for evaluation of community detection methods, too.

In order to evaluate the clustering results more objectively, a large number of external evaluation indices were proposed.
\subsubsection{\label{external}External indices}
Generally speaking, external indices evaluate the clustering result by comparing it with the external supplied true labels. For the sake of clarity, we assume that there are 10 samples with the true labels \textsl{[1, 1, 1, 1, 1, 1, 2, 2, 3, 3]}.\\
\begin{enumerate}[i)]
\item Purity \cite{Manning2008Introduction}: This index maps each cluster obtained by clustering method to the ground-truth one with highest overlap, and tries to calculate the percentage of ``corrected'' clustered samples, leading to the following formula:
\begin{equation}
Purity\left(B,A\right)=\dfrac{1}{n}\sum_{k}\max_{r}\left|B_{k} \cap A_{r}\right|,
\end{equation}
where $n$ is the number of samples, $B=\{B_{1},B_{2},\dots,B_{c}\}$ and $A=\{A_{1},A_{2},\dots,A_{c^*}\}$ are sets of clusters obtained by clustering method and the ground-truth ones, respectively.

For example, if the labels obtained by clustering method are [2, 2, 2, 2, 1, 1, 1, 1, 3, 3], purity will match 2 to \textsl{1}, 1 to \textsl{2}, 3 to \textsl{3}, and the purity value is $0.80$. Noting that the mapping is not  one-to-one, but many-to-one. For example, if the labels obtained by clustering method are [1, 1, 1, 4, 4, 4, 2, 2, 3, 3], purity will match 1 to \textsl{1}, 4 to \textsl{1}, 2 to \textsl{2} and 3 to \textsl{3}. The purity value is 1, which is apparently unreasonable.

Rossetti, et.al \cite{rossetti2016novel} proposed community precision, community recall and F1-measure to evaluate the performance of community detection methods. However, the mapping in the method is not one-to-one either, leading to the same problem that purity has. For example, if the labels obtained by clustering method are [1, 1, 1, 1, 1, 1, 1, 1, 2, 2], the mapping will be: $1\mapsto\textsl{1}, 2\mapsto\textsl{3}$, and the value of F1 is 0.93. On the other hand, if the labels obtained by clustering method are [1, 1, 1, 1, 1, 2, 2, 2, 3, 3], the mapping will be: $1\mapsto\textsl{1}, 2\mapsto\textsl{2}, 3\mapsto\textsl{3}$, and the value of F1 is 0.90. The results are apparently unreasonable.

\item Rand index \cite{Rand1971Objective}: The most important information we can get from clustering result is not the labels, but rather which samples are clustered together and which ones are not, motivating people to define Rand index as follows:\\
\begin{equation}
Rand\left(A,B\right)=\dfrac{2\left(a+d\right)}{n\left(n-1\right)},
\end{equation}
where $n$ is the number of samples, $B=\{B_{1},B_{2},\dots,B_{c}\}$ and $A=\{A_{1},A_{2},\dots,A_{c^*}\}$ are sets of detected clusters and the ground-truth ones, respectively. $a$ is the number of pairs of samples which are in the same cluster of both $A$ and $B$, and $d$ is the number of pairs of samples which are not in the same cluster of both $A$ and $B$.

For example, if the labels obtained by clustering method are [1, 1, 1, 4, 4, 4, 2, 2, 3, 3], the Rand index is 0.8, which is better than purity. However, Rand index does not notice the importance of small clusters in imbalanced data. For example, both of the clustering results [1, 1, 1, 1, 1, 1, 1, 2, 3, 3] and [1, 1, 1, 1, 1, 2, 2, 2, 3, 3] have the same value of Rand index 0.84, which is not informative.
\item Normalized mutual information (NMI, \cite{Danon2005Comparing,Strehl2003Cluster}): NMI derives from entropy in information theory. For a discrete random variable $X$, its entropy is defined as:\\
\begin{equation}
H\left(X\right)=-E\left[\log p\left(x\right)\right]=-\sum_{x}p\left(x\right)
\log p\left(x\right).
\end{equation}
For example, the entropy of the true labels is:
\begin{equation}
H_1=-\left(\frac{6}{10}\log\frac{6}{10}+\frac{2}{10}\log\frac{2}{10}+\frac{2}{10}\log\frac{2}{10}\right)=1.37.
\end{equation}
Similarly, the joint entropy can be defined:\\
\begin{equation}
\begin{split}
H\left(X,Y\right) & =-E\left[\log p\left(x,y\right)\right]\\[4pt] & =-\sum_{x}\sum_{y}p\left(x,y\right)\log p\left(x,y\right),
\end{split}
\end{equation}
and conditional entropy is:
\begin{equation}
\begin{split}
H\left(Y|X\right)&=-E\left[H\left(y|x\right)\right]=-\sum_{x}p\left(x\right)H\left(y|x\right)\\&=-\sum_{x}\sum_{y}p\left(x,y\right)\log p\left(y|x\right).
\end{split}
\end{equation}
Based on the above definitions, the mutual information of two random discrete variables can be defined as:\\
\begin{equation}
I(X;Y) = H(X)-H(X|Y),
\end{equation}
which measures the mutual dependence between $X$ and $Y$.

The relations among entropy, joint entropy, conditional entropy and mutual information are shown in Fig. \ref{Fig:01}.
 \begin{figure}[!t]
	\centering
	\includegraphics[height=4cm, width=7cm]{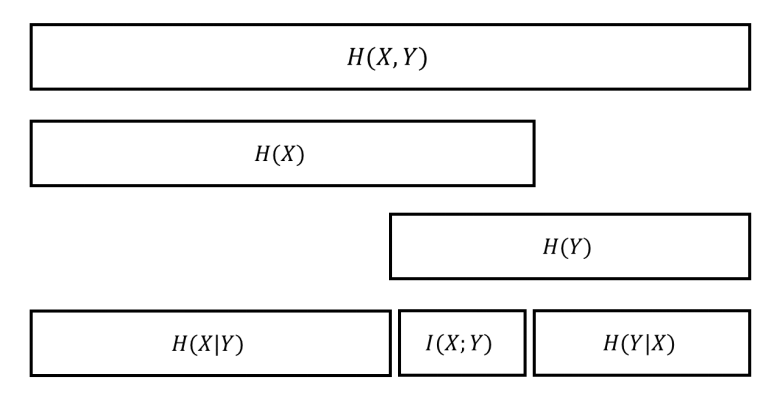}
	\caption{Relations among entropy, joint entropy, conditional entropy and mutual information}\label{Fig:01}
\end{figure}

The mutual information between ground-truth labels and the labels obtained by clustering method can be used to evaluate the quality of clustering result. However, the range of mutual information is not between 0 and 1. For example, the mutual information $I(X;X)$ is actually $H(X)$, and can be any non-negative real value, making the evaluation result nonintuitive and hard to explain. Normalized mutual information (NMI) is thus proposed as follows:
\begin{equation}
NMI\left(X,Y\right)=\dfrac{2I\left(X;Y\right)}{H\left(X\right)+H\left(Y\right)}
\end{equation}
with value ranging from 0 to 1. Now NMI has become the most widely used index for evaluation of clustering methods and is also for evaluation of community detection methods. NMI notices the importance of small clusters to some degree, but does not solve the problem completely.  For example, the NMI values of two computed clustering results [1, 1, 1, 1, 1, 1, 1, 2, 3, 3] and [1, 1, 1, 1, 1, 2, 2, 2, 3, 3] are 0.76 and 0.77, respectively, which are in line with our expectations. However, the NMI values of [1, 1, 1, 1, 1, 1, 2, 2, 2, 3] and [1, 1, 1, 1, 1, 2, 2, 2, 3, 3] are 0.82 and 0.77 respectively, which are counter-intuitive.

In addition, the probabilities in NMI are approximated by frequency, resulting in finite size effect and preferring large number of partitions, and rNMI was proposed to overcome this issue \cite{zhang2015evaluating}. However, rNMI has reverse finite size effect also due to finite size of network, making it prefer small number of partitions, and cNMI was proposed to remove both finite size effect of NMI and reverse finite size effect of rNMI \cite{Lai2016A}. We use Fig. \ref{Fig:02}\cite{Lai2016A} to better demonstrate the above conclusions: we calculate NMI, rNMI and cNMI between two partitions $A$ and $B$, where $A$ is the ground-truth community label in LFR network, and $B$ is obtained from $A$ by flipping the labels of a fraction of nodes. One can observe that: 1) The values of NMI between two independent partitions are much larger than zero due to finite size effect; 2) The values of rNMI between two identical partitions are not one due to reverse finite size effect; 3) cNMI can handle both finite size effect and reverse finite size effect, but the lines are not straight, indicating that the decline of cNMI is not proportional to the fraction of flipping labels, which is counter-intuitive.
 \begin{figure*}
 \centering
\includegraphics[height=5cm, width=17cm]{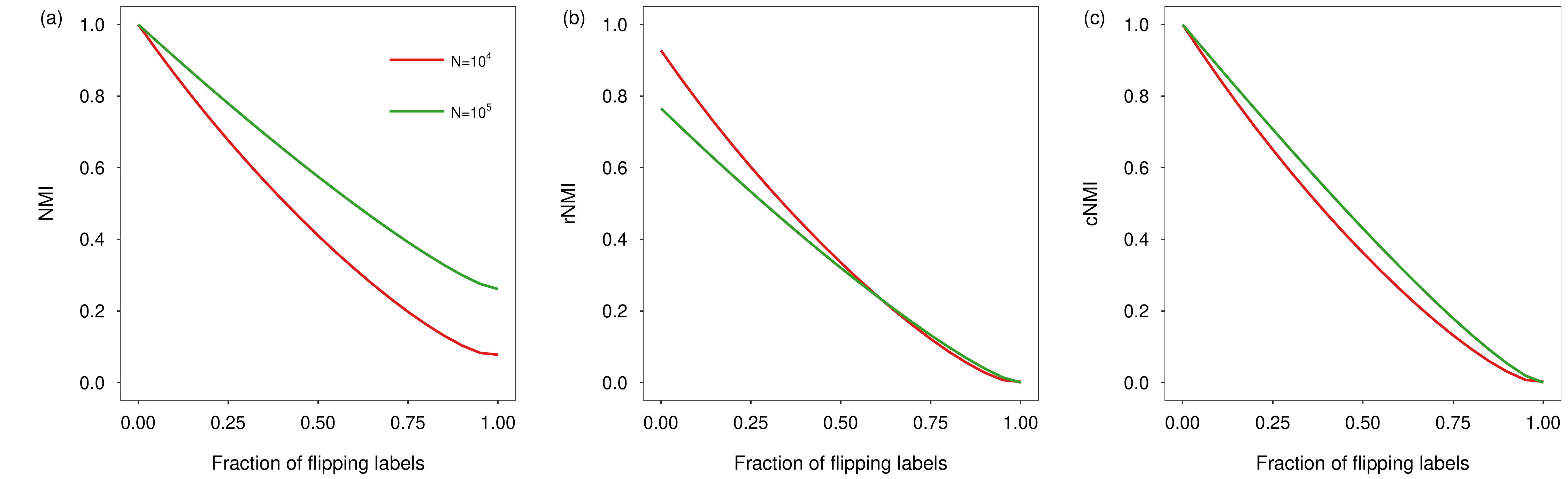}
\caption{Normalized mutual information (NMI), relative normalized mutual information (rNMI) and corrected normalized mutual information (cNMI) between two partitions $A$ and $B$, where $B$ is obtained from $A$ by flipping the labels of a fraction of nodes. NMI of two independent partitions is larger than 0, rNMI of two identical partitions is smaller than 1, and cNMI can handle these issues. However the lines of cNMI are not straight.}\label{Fig:02}
\end{figure*}

Finally, it is worth mentioning that neither rNMI nor cNMI can solve the problem of ignoring importance of small communities. For example, for the clustering results [1, 1, 1, 1, 1, 1, 2, 2, 2, 3] and [1, 1, 1, 1, 1, 2, 2, 2, 3, 3], the values of rNMI are 0.57 and 0.47, respectively, and those of cNMI are 0.76 and 0.64, respectively. The results are still counter-intuitive.

More discussions of random effect on clustering similarity can be found in \cite{Gates2017The}.
\end{enumerate}

Besides the problems discussed above, the indices of clustering validity in unsupervised learning cannot be used to evaluate specific community of interest. However, under many circumstances, we prefer the methods who can detect specific communities or structures in the network even if their overall performance is not good.

In the next section, we will briefly describe how to evaluate classification results in supervised learning.
\begin{table}
\caption{Confusion matrix of binary classification}
\centering
\begin{tabular}{ccccc}
	\hline\hline
	\quad & \quad & \multicolumn{3}{c}{Predicted class} \\
	\hline
	\quad & \quad & \quad & 1 & 0\\
	\multirow{2}{0.1cm}{\rotatebox{90}{Actual}} & \multirow{2}{0.4cm}{\rotatebox{90}{class}} & 1 & \hspace{5mm}True Positive (TP)\hspace{5mm} & False Negative (FN)\\
	& & 0& \hspace{5mm}False Positive (FP)\hspace{5mm} & True Negative (TN)\\
	\hline\hline
\end{tabular}\label{Tab:01}
\end{table}
\subsection{\label{classification}Indices of classification validity in supervised learning}
We start with binary classification, and the confusion matrix, which displays the classification results intuitively, can be defined as Table \ref{Tab:01}. Based on confusion matrix, one can define accuracy, sensitivity, specificity, precision, recall and F-score as follows\cite{Metz1978Basic,BIA0001114,Galton1892,van1979information}:
\begin{eqnarray}
Accuracy & = & \dfrac{TP+TN}{TP+FP+FN+TN} \\[3mm]
Sensitivity & = & \dfrac{TP}{TP+FN} \\[3mm]
Specificity & = & \dfrac{TN}{TN+FP} \\[3mm]
Precision & = & \dfrac{TP}{TP+FP} \\[3mm]
Recall & = & \dfrac{TP}{TP+FN}\\[3mm]
\hspace{-15mm}F-score & \hspace{-5mm}= & \hspace{-5mm}\left(1+\beta ^{2}\right) \cdot \dfrac{precision \cdot recall}{\left(\beta ^{2} \cdot precision \right) + recall}
\end{eqnarray}

Accuracy is the percentage of correctly classified samples, and does not consider the significance of small class. Sensitivity/recall (specificity) calculates how good a method is at detecting the positives (avoiding false positives), and can be maximized by labeling all of the samples with positive (negative). Precision calculates the fraction of samples labeled with positive that are really positives, and can be maximized by only labeling one positive sample with positive. In summary, the above metrics may be misleading, and F-score is proposed to solve the problems by combining precision and recall instead of just using one.

Furthermore, kappa index\cite{Galton1892} considers the influence of random effects, and is defined as:
\begin{equation}
kappa=\dfrac{accuracy-expected\,\, accuracy}{1-expected\,\, accuracy}
\end{equation}

Kappa index can overcome the challenge of ignoring small class mentioned in Sect.\ref{external}: {\it External indices}, i.e., if the true labels are \textsl{[1, 1, 1, 1, 1, 1, 2, 2, 3, 3]}, the kappa indices for the two classification results [1, 1, 1, 1, 1, 1, 2, 2, 2, 3] and [1, 1, 1, 1, 1, 2, 2, 2, 3, 3] are 0.82 and 0.83, respectively. This meets our expectations.


The metrics can be naturally extended for evaluation of multi-class classification problems. Noting that the indices mentioned above can be used to evaluate specific class of interest.

In next section, we connect the performance evaluation problem of clustering to classification through integer linear programming. Then we use kappa index for overall performance evaluation of community detection methods, and F-score for evaluation of detecting a single community of interest.

\section{Model Description}\label{model}
In this section, we map the computed labels to the ground-truth ones through integer linear programming. The mapping is one-to-one.

We assume without loss of generality that $c \leqslant c^*$, and introduce the following decision variables:
\begin{equation}
x_{ij} = \left\{\begin{array}{l}\vspace{2mm}
                             1, \mbox{\hspace{5mm}computed label $i$ is mapped to ground truth $j$.}\\
                             0, \mbox{\hspace{5mm}otherwise.}
                             \end{array}
                             \right.
\end{equation}
The cost of the mapping is $l_{ij} = |B_i\cup A_j|-|B_i\cap A_j|$, where $A=\{A_{1},A_{2},\dots,A_{c^*}\}$ and $B=\{B_{1},B_{2},\dots,B_{c}\}$ are set of ground-truth communities and set of computed communities, respectively.
The complete integer linear programming model is:
\begin{equation}\label{eq:02}
\begin{array}{rl}
 \min & z = \displaystyle\sum\limits_{i=1}^c\sum\limits_{j=1}^{c^*}l_{ij}x_{ij}\\[7mm]
 s.t. & \left\{
               \begin{array}{l}
                 \displaystyle\sum\limits_{i=1}^cx_{ij}\leq1, \, j=1,\, 2,\, \cdots,\, c^*.\\[7mm]
                 \displaystyle\sum\limits_{j=1}^{c^*}x_{ij} = 1, \, i=1,\, 2,\, \cdots,\, c.\\[7mm]
                 x_{ij} = 0\ or\ 1.
               \end{array}
        \right.
\end{array}
\end{equation}
Actually, it can be reformulated as standard assignment problem by creating the cost matrix as $L^\prime = (l_{ij}^\prime)_{c^*\times c^*}$ such that:


\begin{equation}
l_{ij}^\prime = \left\{\begin{array}{l}\vspace{2mm}
                             l_{ij}, \mbox{\hspace{5mm}if $i\leq c$}\\
                             0, \mbox{\hspace{7mm}if $i > c$,}
                             \end{array}
                             \right.
\end{equation}
i.e.,
\begin{equation}
L^\prime=\begin {array}{rl}
                \begin{bmatrix}
                  l_{11} & l_{12} & \cdots & \cdots & \cdots & \cdots & l_{1c^*} \\
                  l_{21} & l_{22} & \cdots & \cdots & \cdots & \cdots & l_{2c^*} \\
                  \cdots & \cdots & \cdots & \cdots & \cdots & \cdots & \cdots \\\vspace{2mm}
                  l_{c1} & l_{c2} & \cdots & \cdots & \cdots & \cdots & l_{cc^*} \\
                  0   &   0    & \cdots & \cdots & \cdots & \cdots & 0      \\
                  \cdots & \cdots & \cdots & \cdots & \cdots & \cdots & \cdots \\
                  0   &   0    & \cdots & \cdots & \cdots & \cdots & 0      \\
                \end{bmatrix}
                &\hspace{-5mm}\begin {array}{c}
                               \vspace{-50mm}\hspace{2mm}c^*\times c^*.
                              \end {array}
         \end {array}
\end{equation}
And the model is updated:
\begin{equation}\label{eq:03}
\begin{array}{rl}
 \min & z = \displaystyle\sum\limits_{i=1}^{c^*}\sum\limits_{j=1}^{c^*}l^\prime_{ij}x_{ij}\\[7mm]
 s.t. & \left\{
               \begin{array}{l}
                 \displaystyle\sum\limits_{i=1}^{c^*}x_{ij}=1, \, j=1,\,2,\,\cdots,\,c^*.\\[7mm]
                 \displaystyle\sum\limits_{j=1}^{c^*}x_{ij} = 1, \, i=1,\,2,\,\cdots,\,c^*.\\[7mm]
                 x_{ij} = 0\ or\ 1,
               \end{array}
        \right.
\end{array}
\end{equation}
which can be solved by Hungarian method.

The following theorem proves that the optimal solution of (\ref{eq:02}) is guaranteed by solving (\ref{eq:03}).
\begin{theorem}
 If $c\leqslant c^*$ and $X^{*\prime} = (x_{ij}^{*\prime})_{c^*\times c^*}$ is the optimal solution of (\ref{eq:03}), then $X^* = (x^*_{ij})_{c\times c^*}$ is the optimal solution of (\ref{eq:02}),   where $x_{ij}^{*}=x_{ij}^{*\prime},\, i = 1, 2, \cdots, c,\, j = 1, 2, \cdots, c^*.$
\end{theorem}

\begin{IEEEproof}
Obvious if $c = c^*.$

Otherwise, since $X^{*\prime}$ is the optimal solution of problem (\ref{eq:03}), one has: $\forall\ \mbox{matrix}\ X \mbox{ of size}\ c\times c^*$, if $X$ is the feasible solution of problem (\ref{eq:02}), one can create $X^\prime$ as follows:
\begin{equation}
x_{ij}^\prime = \left\{\begin{array}{l}\vspace{2mm}
                             x_{ij}, \mbox{\hspace{5mm}if $i\leq c$,}\\\vspace{2mm}
                             0, \mbox{\hspace{7mm}if $i > c$, and $\sum\limits_{i=1}^cx_{ij}=1$}\\\vspace{2mm}
                             1, \mbox{\hspace{7mm}if $i > c$, and $\sum\limits_{i=1}^cx_{ij}=0,$ $\sum\limits_{t=1}^{j-1}x_{it}=0.$}\\
                             0, \mbox{\hspace{7mm}if $i > c$, and $\sum\limits_{i=1}^cx_{ij}=0,$ $\sum\limits_{t=1}^{j-1}x_{it}=1.$}
                             \end{array}
                             \right.
\end{equation}
$X^\prime$ is obviously the feasible solution of problem (\ref{eq:03}), hence

\begin{equation}
\sum_{i=1}^{c^*}\sum_{j=1}^{c^*} l_{ij}^\prime x_{ij}^{*\prime}\leq\sum_{i=1}^{c^*}\sum_{j=1}^{c^*} l^\prime_{ij}x_{ij}^\prime.
\end{equation}

In other words,
\begin{equation}
    \sum_{i=1}^c\sum_{j=1}^{c^*} l_{ij} x^*_{ij}\leq\sum_{i=1}^c\sum_{j=1}^{c^*} l_{ij}x_{ij}
\end{equation}
based on the definition of $l_{ij}^\prime$, meaning that $X^*$ is the optimal solution of (\ref{eq:02}).
\end{IEEEproof}

For example, if the ground-truth labels and the computed labels are \textsl{[1, 1, 1, 1, 4, 4, 2, 2, 3, 3]} and [3, 3, 3, 3, 3, 3, 3, 1, 1, 2], respectively, then $l_{11} = 6,$ $l_{12} = 3-1 = 2,$ $l_{13} = 3-1 = 2,$ $l_{14} = 4$, $l_{21} = 5,$ $l_{22} = 3,$ $l_{23} = 1$, $l_{24} = 3,$ $l_{31} = 7-4 = 3,$ $l_{32} = 8-1 = 7$, $l_{33} = 3$, $l_{34} = 7-2 = 5$ and the resulting map is: $1\mapsto\textsl{2},\, 2\mapsto\textsl{3},\, 3\mapsto\textsl{1}$, making it possible to use indices in the field of supervised learning, such as kappa index and F-score for evaluation of community detection methods. It is worth mentioning that the method also applies to overlapping community structures detection. For example, if the ground truth labels are \textsl{[1, 1, 1, 1, (1,2), (1,2), 2, 2, 3, 3]}, where nodes $5$ and $6$ have multiple labels, and the computed labels are [3, 3, 3, 3, 1, (1,3), 1, 1, 2, 2], then the costs of mapping are: $l_{11} = 8-2 = 6,$ $l_{12} = 4-4 = 0,$ $l_{13} = 6-0 = 6$, $l_{21} = 8,$ $l_{22} = 6,$ $l_{23} = 2-2 = 0,$ $l_{31} = 6-5 = 1,$ $l_{32} = 8-1 = 7,$ $l_{33} = 7$, and the resulting mapping is $1\mapsto\textsl{2},\, 2\mapsto\textsl{3},\, 3\mapsto\textsl{1}$. However, how to evaluate overlapping community detection methods is not the focus of this paper, and we will leave it for further study.

The time complexity of obtaining the cost matrix in the integer linear programming is $O(cc^* )$, where $c$ and $c^*$ are the numbers of computed communities and the ground-truth ones, respectively. The integer linear programming model can be solved by Hungarian method (also known as Kuhn-Munkres algorithm), whose time complexity is $O(c^2 c^*)$ with $c\leqslant c^*$\footnote{http://zafar.cc/2017/7/19/hungarian-algorithm/}.

Actually, the numbers of communities do not increase proportionally with growing network sizes. For example, the numbers of communities in LFR networks used in the paper are about $0.004n$, where $n$ is the network size, as is shown in Fig. \ref{Fig:09}; the numbers of communities in real networks used in our paper and ref. \cite{harenberg2014community} are also very small compared with the network sizes, as is summarized in Table \ref{Tab:02}. Note that the networks used in ref. \cite{harenberg2014community} are obtained after preprocessing the data from the Stanford Large Network Dataset Collection
(SNAP)\footnote{http://snap.stanford.edu/} and are not released. In addition, effective algorithms for large scale assignment problems were proposed \cite{Hong2016}. Hence, the proposed method can be applied to large scale networks.
\begin{table}[]
\centering
\renewcommand{\arraystretch}{2.5}
\caption{Summary of nine real networks commonly used in community structures detection research.}
\begin{tabular}{c | ccc}\hline\hline
 \backslashbox{\footnotesize Networks}{\footnotesize Parameters} & Num vertices             & Num Edges                 & Num communities        \\\hline
Email-Eu-core & 1005 & 25571 & 42 \\
Orkut & 1247 & 15511 & 43 \\
Political blogs & 1490 & 19025 & 2  \\
Amazon                    & 1644                     & 5022                      & 131                    \\
Cora            & 2708 & 5429  & 7 \\
Youtube                   & 3088                     & 6695                      & 595                    \\
CiteSeer        & 3312 & 4732  & 6  \\
DBLP                      & 5831                     & 18733                     & 676                    \\
Livejournal               & 17969                    & 434075                    & 668                    \\
\hline\hline
\end{tabular}\label{Tab:02}
\end{table}
\section{Experimental Results}\label{experiments}
In this section, we will firstly demonstrate the performance advantages of using kappa and F-score compared with NMI, then systematically evaluate several popular community detection methods on both synthetic benchmark networks and real networks using our method.
\subsection{Description of Synthetic Networks}
The Lancichinetti-Fortunato-Radicchi (LFR) benchmark network model\cite{lancichinetti2008benchmark} is the most popular used one for evaluation of community detection algorithms because it is a good proxy for real networks. The LFR network is generated by several parameters, including number of total nodes $n$, average degree $k$, maximal degree $k_{max}$, minimal community size $s_{min}$ and maximal community size $s_{max}$, exponent of power-law distributions of
nodes degree $\gamma$ and community size $\beta$, respectively, and a mixing ratio of external links $\mu$. Specifically, the higher the mixing parameter $\mu$ of a network is, the more difficult it is to detect the community.
\subsection{Description of Real Networks}
Four real networks, including email-Eu-core network \cite{yin2017local,leskovec2007graph}, Political blogs network \cite{adamic2005political}, CiteSeer network \cite{lu2003link,sen2008collective} and Cora network \cite{lu2003link,sen2008collective} are used for our experiments. Details are described below:
\begin{enumerate}
\item email-Eu-core network: This data set is the e-mail network of members in a European research institute. The members are represented as nodes, and a directed edge $(i,j)$ means member $i$ sent to member $j$ at least one email. There are 1005 nodes and 25571 edges. Each member belongs to one of 42 departments at the institute.
\item Political blogs network: This data set is a network of hyperlinks between weblogs on US politics. There are 1490 nodes and 19025 edges. Each node is either conservative or liberal.
\item CiteSeer network: This data set is a citation network. The nodes of the network are publications and the directed edges denote citations. There are 3312 nodes and 4732 edges. Each publication belongs to one of six classes, including Agents, Artificial Intelligence, Database, Human Computer Interaction, Information Retrieval, and Machine learning.
\item Cora network: This data set is also a citation network. There are 2708 scientific publications and 5429 edges. Each publication belongs to one of seven classes, including Case-based Reasoning, Genetic Algorithms, Neural Networks, Probabilistic Methods, Reinforcement Learning, Rule Learning, and Theory.
\end{enumerate}

\begin{figure*}[ht]
\centering
\includegraphics[height=5cm,width=17cm]{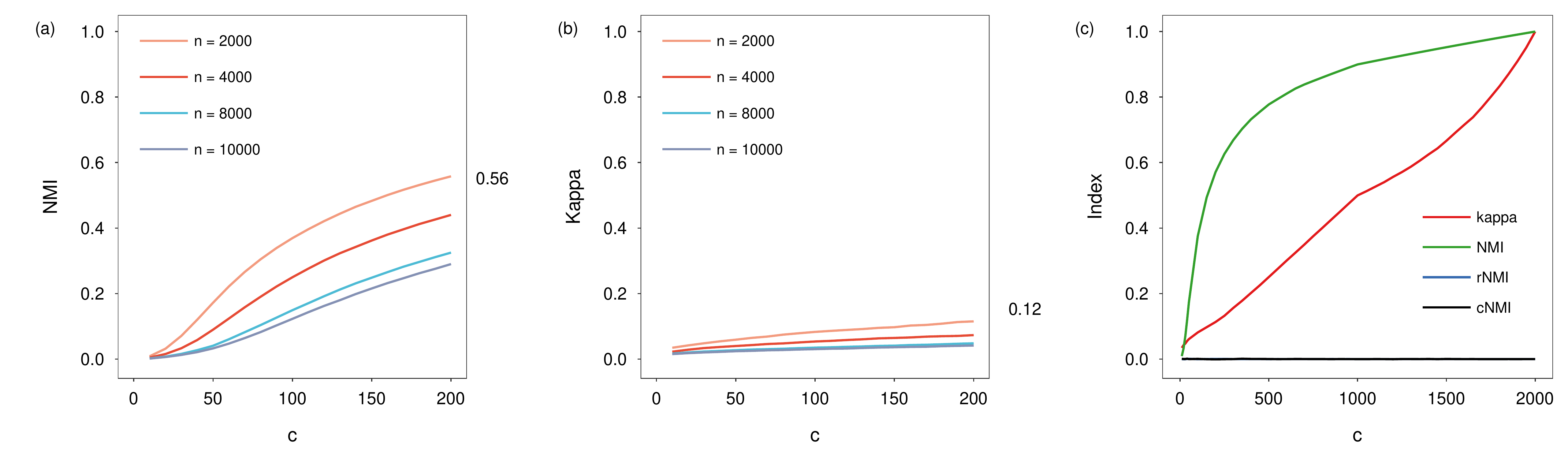}
\caption{Comparison between two independent random partitions $A$ and $B$ using NMI, kappa index, rNMI and cNMI. The community number $c$ ranges from 10 to 200, and the expected community size is $\displaystyle\frac{n}{c}$, where $n$ is the number of nodes. The results are averages of ten trials.}\label{Fig:03}
\end{figure*}
\begin{figure*}[ht]
\centering
\includegraphics[height=5cm,width=17cm]{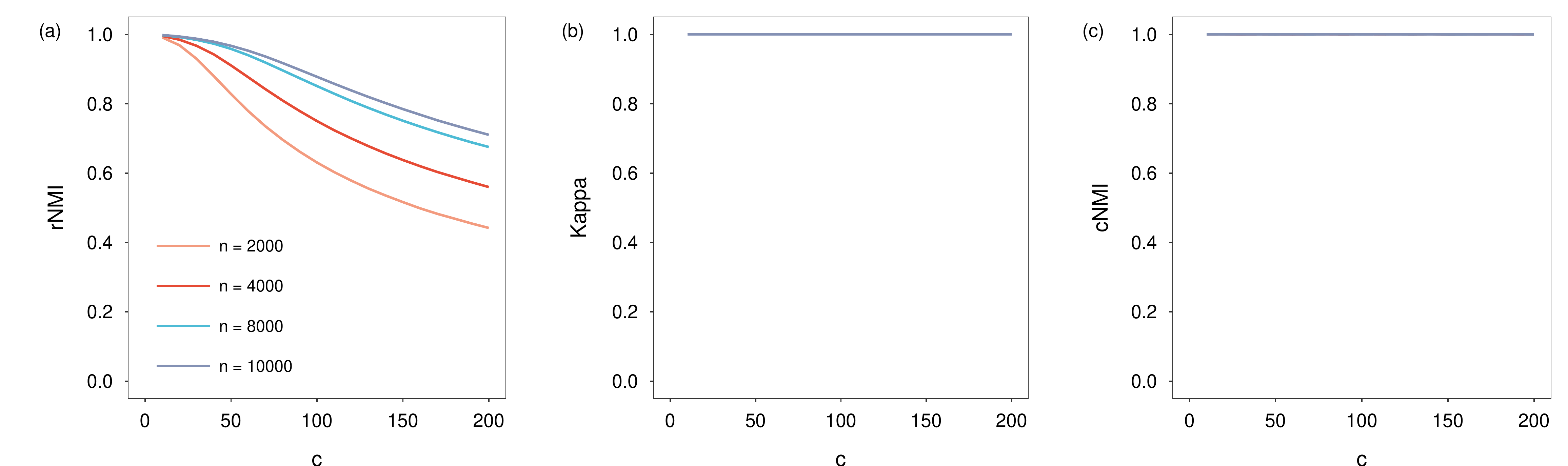}
\caption{rNMI, kappa index and cNMI of $A$ against itself. The community number $c$ ranges from 10 to 200, and the expected community size is $\displaystyle\frac{n}{c}$, where $n$ is the number of nodes. The results are averages of ten trials.}\label{Fig:04}
\end{figure*}

\subsection{Advantages of Kappa Index compared with NMI and rNMI}
As we have mentioned above, NMI has finite size effect, and rNMI has reverse finite size effect.
Finite size effect can be demonstrated by calculating NMI between two independent random partitions $A$ and $B$, and reverse finite size effect can be demonstrated by calculating rNMI of $A$ against itself. The results and the advantages of kappa index are demonstrated in Fig. \ref{Fig:03} (a), (b) and Fig. \ref{Fig:04} (a), (b). One can observe that: 1) NMI between two independent random partitions is much larger than zero, and is farther away from zero for partitions with larger number of communities due to finite size effect. This phenomenon is more evident in smaller networks. 2) rNMI of a partition against itself is not one, and is farther away from one for partitions with larger number of communities due to reverse finite size effect. This phenomenon is also more evident in smaller networks. 3) The values of kappa are not strictly constant in Fig. \ref{Fig:03} (b), especially for large numbers of communities.

Actually, the index value between two random and independent partitions $A$ and $B$ depends on the network size $n$ and the community number $c$, and should not be zero, whatever the index is.
For example, there are three samples with two clusters, possible partitions include $P = \{(1,\,2,\,2),\, (1,\,1,\,2),\,(1,\,2,\,1),\,(2,\,1,\,1),\,(2,\,1,\,2),\,$ $(2,\,2,\,1)\}$, and none of the values between $A$ and $B$ that are randomly selected from the set $P$ are zero. In extreme circumstances where $c = n$, the reasonable value should be 1.

\begin{figure}[ht]
\centering
\includegraphics[height=5cm,width=6cm]{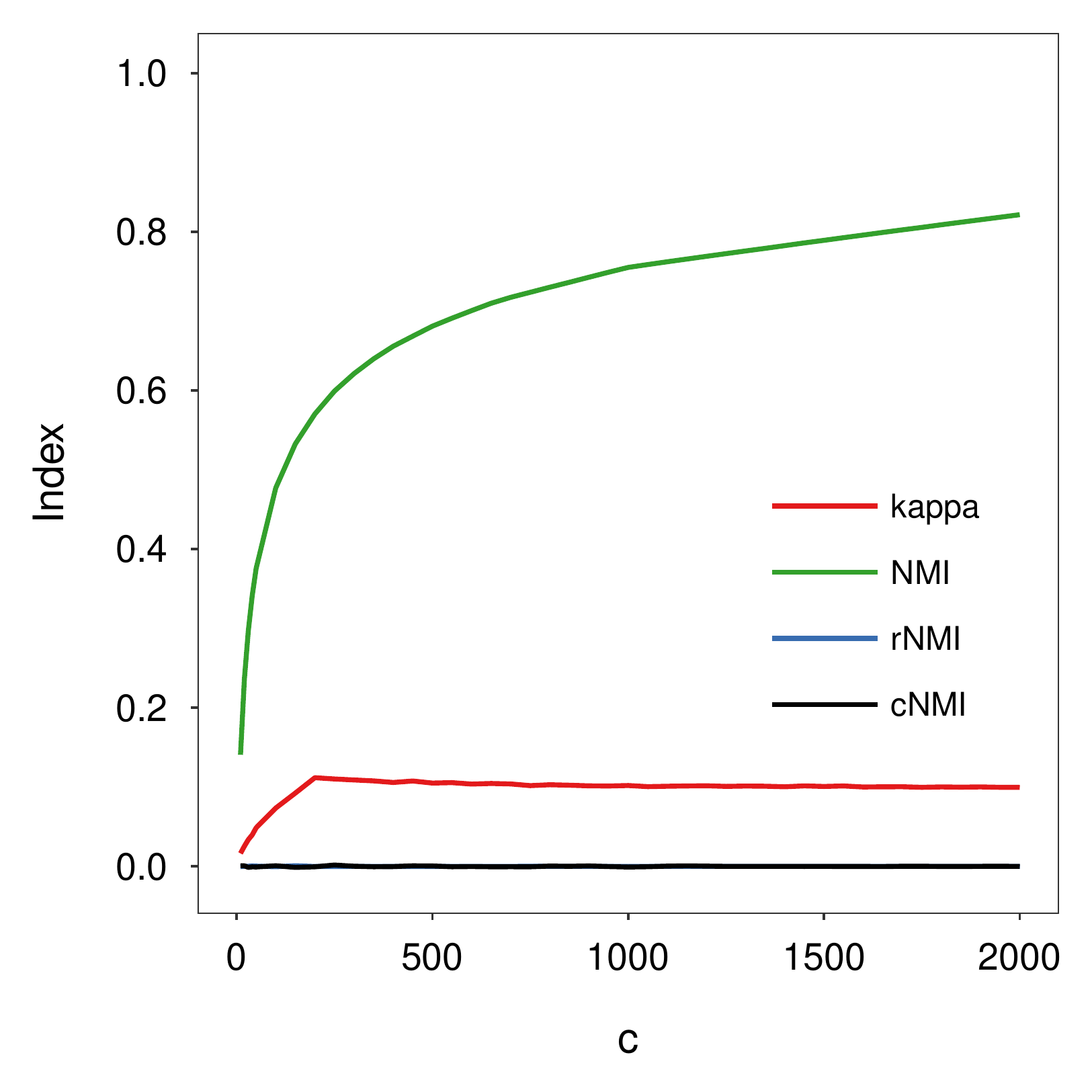}
\caption{
Comparison between two independent random partitions $A$ and $B$ using kappa, NMI, rNMI and cNMI. The network size $n$ is 2000. $A$ always has $200$ communities, and the expected community size is $10$. $B$ has a varying number of communities $c$, and the expected community size is $\frac{2000}{c}$.}
\label{Fig:12}
\end{figure}

\begin{figure*}[ht]
\centering
\includegraphics[height=5cm,width=13cm]{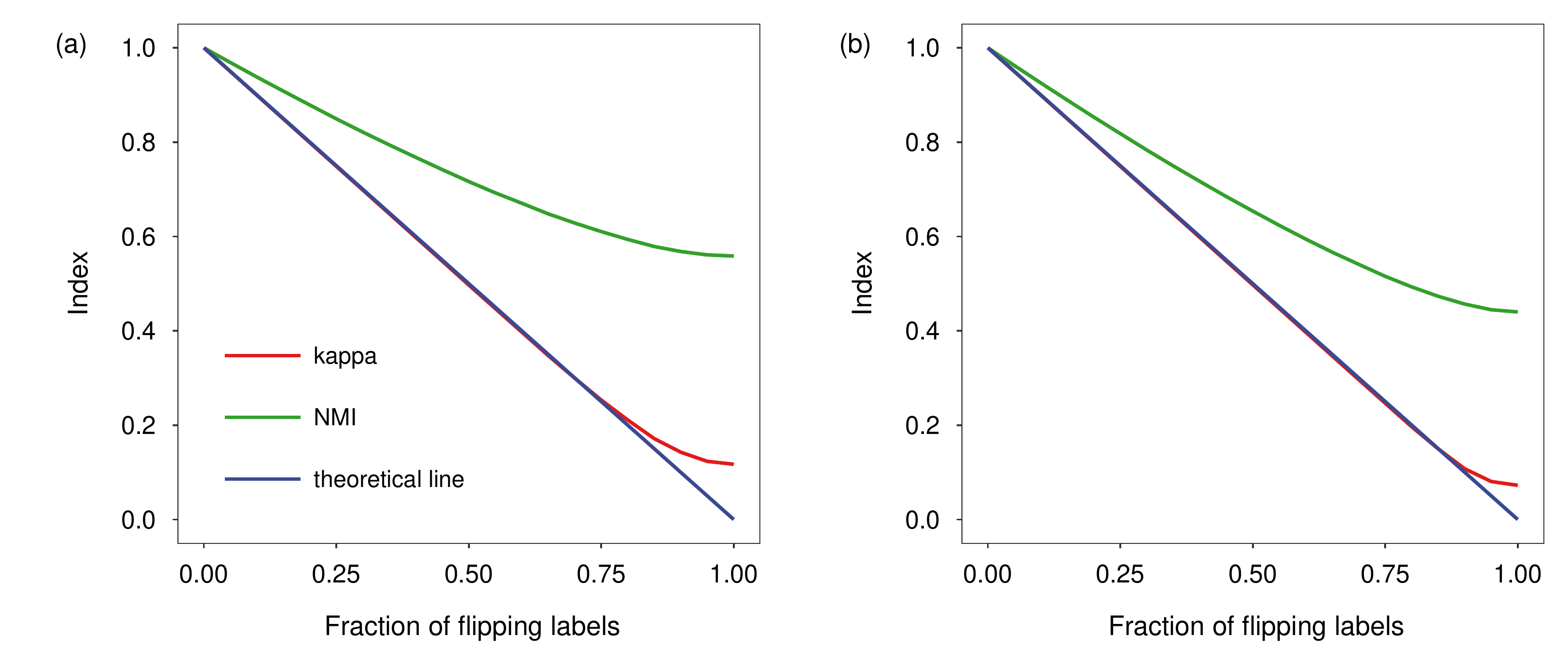}
\caption{Kappa index and NMI of partition $A$ versus $B$, where $A$ is (a) uniformly random partition with network size $n=2000$ and community number $c=200$; (b) uniformly random partition with network size $n=4000$ and community number $c=200$, and $B$ is obtained through randomly flipping the labels of a fraction of $A$.}\label{Fig:13}
\end{figure*}
\begin{figure*}[ht!]
\centering
\includegraphics[height=50mm,width=150mm]{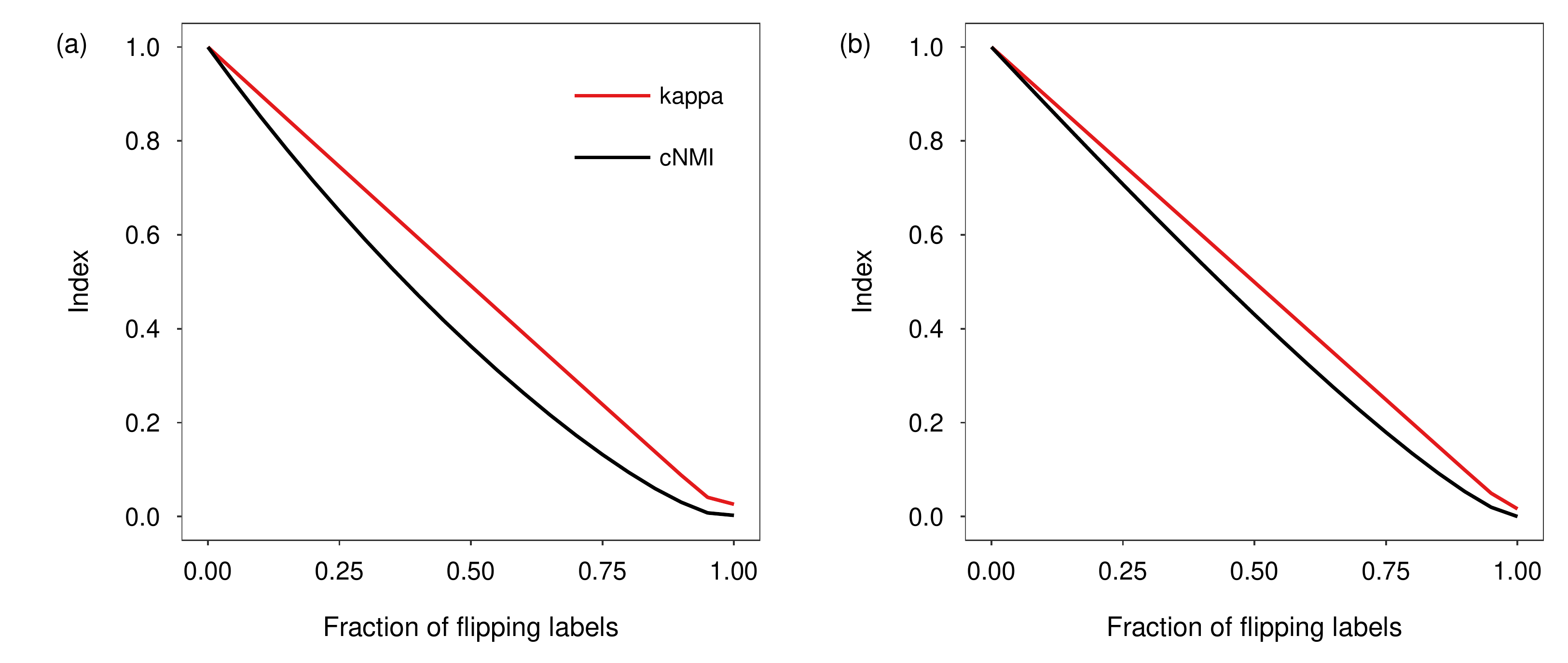}
\caption{cNMI and kappa index of two partitions $A$ and $B$. $A$ contains the ground-truth labels of LFR networks with (a) $n = 10000$ and (b) with $n = 100000$, respectively. $B$ is obtained through randomly flipping the labels of a fraction of $A$. The results are averages of ten trials. The line of kappa index is straight.}\label{Fig:05}
\end{figure*}

Fig. \ref{Fig:03} (c) calculates the values of kappa, NMI, rNMI and cNMI between two random and independent partitions $A$ and $B$ of networks with $n = 2000$. Fig. \ref{Fig:03} (a), (b) are parts of Fig. \ref{Fig:03} (c). The results are averages over ten trials. One can observe that: 1) The values of kappa increase near linearly with growing community numbers. 2) The gap between NMI and kappa is very vast due to finite size effect of NMI. 3) The values of rNMI and cNMI are always zero, and are apparently unreasonable.

In addition, we also calculate the values of kappa, NMI, rNMI and cNMI between two random and independent partitions $A$ and $B$ of networks with $n = 2000$, where $A$ has $200$ communities and $B$ has a varying number of communities. The results are averages of ten trials and are shown in Fig. \ref{Fig:12}. One can observe that: 1) NMI increases gradually with growing community numbers. 2) Kappa achieves its maximum value at $c = 200$, where $B$ is most compatible with $A$ since they have the same community number. 3) The values of rNMI and cNMI are still zero, and are apparently unreasonable.

To further observe the behaviors of kappa and NMI, we calculate the values of partition $A$ versus $B$, where $A$ is (a) that in Fig. \ref{Fig:03} with network size $n=2000$ and community number $c=200$; (b) that in Fig. \ref{Fig:03} with network size $n=4000$ and community number $c=200$, and $B$ is obtained through randomly flipping the labels of a fraction of $A$, as is shown in Fig. \ref{Fig:13}. 
From this figure, one can see that: 1) The gap between NMI and the theoretical line is very vast. 2) Generally, kappa is well consistent with theoretical values, except when the flipping fraction is considerable large.

In summary, 1) The results of the proposed method are more reasonable. 2) The behaviors of NMI, rNMI and cNMI cannot meet our expectations, and are not suitable for evaluating community detection methods.  More discussions about cNMI can be found in the next subsection.

\subsection{Advantages of Kappa Index compared with cNMI}
Actually, the discussions on finite size effect and reverse finite size effect are not new, and cNMI has been proposed to handle these issues, as we can be see from Fig. \ref{Fig:03} (c) and Fig. \ref{Fig:04} (c). Except for the unreasonable results in Fig. \ref{Fig:03} (c), there are other problems with cNMI, i.e., violating the proportionality assumption, as is shown in Fig. \ref{Fig:05}. We calculate cNMI and kappa index between two partitions $A$ and $B$, where $A$ contains the ground-truth labels of LFR networks, and $B$ is obtained through randomly flipping the community labels of a fraction of nodes in $A$. One can conclude that: 1) The slope of cNMI keeps decreasing as the similarity of $A$ and $B$ is decreased, indicating that cNMI is not proportionally decreased. 2) The line of kappa index is straight, satisfying proportionality assumption.

In addition, all of NMI, rNMI and cNMI suffer from the problem of ignoring importance of small communities.

In summary, kappa index is advantageous over NMI, rNMI and cNMI.
\subsection{Advantages of F-Score compared with NMI}
\begin{figure}[ht]
\includegraphics[height=30mm,width=90mm]{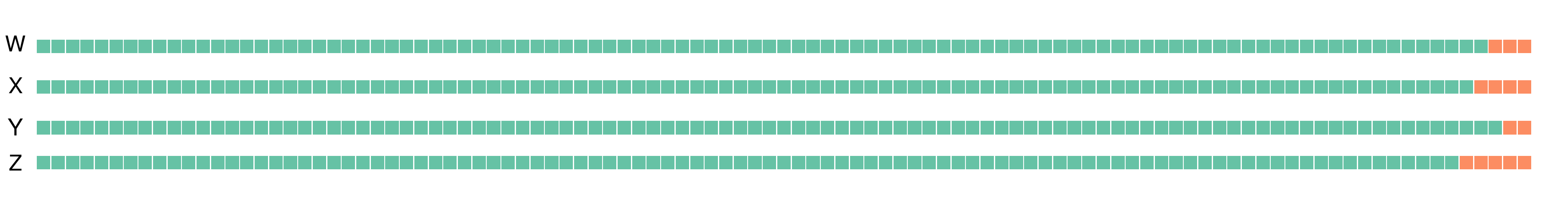}
\caption{Two communities of 103 nodes: $W$ are the ground-truth labels with 100 green ones and 3 orange ones. $X$, $Y$ and $Z$ are the computed labels.}\label{Fig:07}
\end{figure}

Even if ignoring the above discussions, NMI-type metrics cannot be used to evaluate the quality of a single community of interest. For example, given 103 nodes with ground-truth labels $W$, and the computed labels $X$, $Y$ and $Z$, as is shown in Fig. \ref{Fig:07}, the results of NMI, rNMI, cNMI and kappa index are similar, i.e., $\mbox{NMI}(W,X)>\mbox{NMI}(W,Y)>\mbox{NMI}(W,Z),$ $\mbox{rNMI}(W,X)>\mbox{rNMI}(W,Y)>\mbox{rNMI}(W,Z),$ $\mbox{cNMI}(W,X)>\mbox{cNMI}(W,Y)>\mbox{cNMI}(W,Z),$ $\mbox{kappa}(W,X)>\mbox{kappa}(W,Y)>\mbox{kappa}(W,Z).$ These results cannot reveal the methods' ability to detect specific community of interest. Thus we here use F-score to evaluate the two communities separately. The results are $\mbox{F}(W,X)>\mbox{F}(W,Y)>\mbox{F}(W,Z)$ for class orange, and $\mbox{F}(W,Y)>\mbox{F}(W,X)>\mbox{F}(W,Z)$ for class green.

\subsection{Evaluation of Community Structures Detection Methods using Kappa Index}
\begin{figure*}[ht]
\centering
\includegraphics[height=50mm,width=140mm]{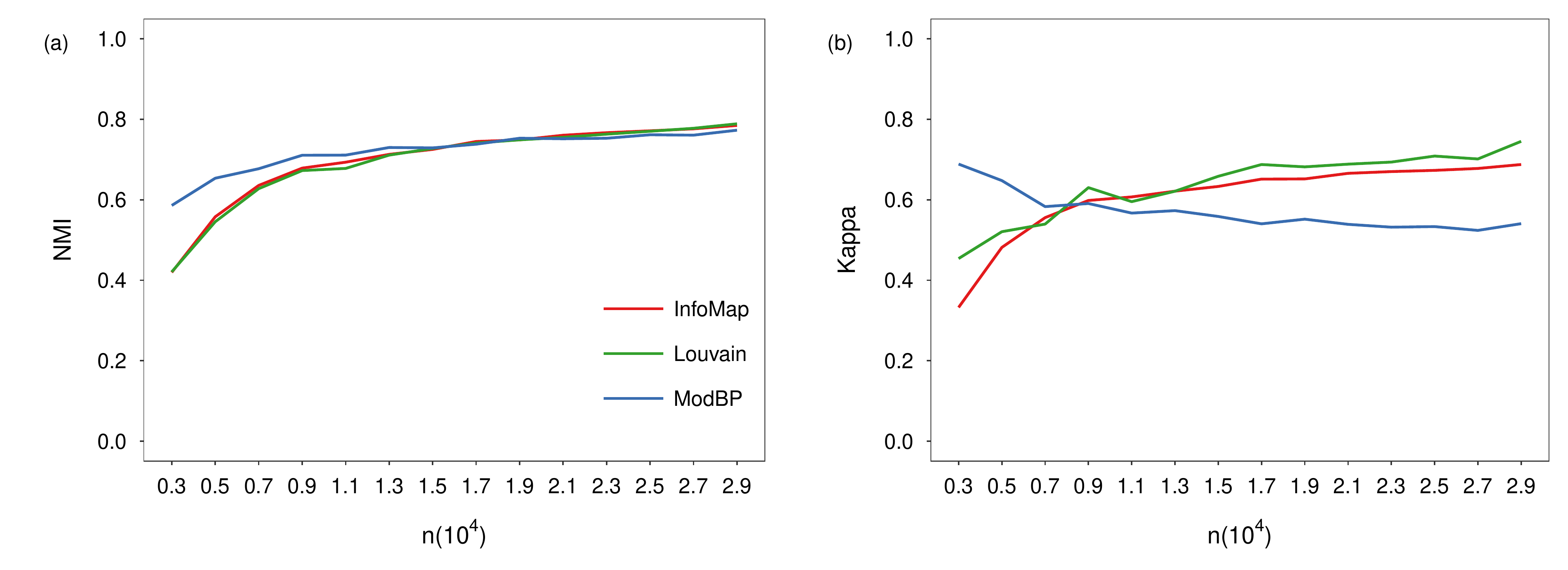}
\caption{NMI and kappa index of three
methods, InfoMap, Louvain and Modularity BP on LFR benchmarks with different network sizes. The parameters of networks are: average degree $k = 8$, mixing parameter $\mu = 0.45$, exponent of degree distribution $\gamma = 2$, exponent of community size distribution $\beta = 1$, maximum degree is $50$, and community sizes range
from $200$ to $400$.}\label{Fig:06}
\end{figure*}

In this subsection, we use normalized mutual information (NMI) and kappa index to rank three methods, InfoMap\cite{rosvall2007maps}, Louvain\cite{blondel2008fast} and Modularity BP\cite{zhang2014scalable} on LFR benchmark networks. The parameters of networks are: average degree $k = 8$, mixing parameter $\mu = 0.45$, maximum degree is $50$, community sizes range
from $200$ to $400$, exponent of degree distribution is $\gamma = 2$, and exponent of community size distribution is $\beta = 1$. The results are averages over ten trials and are shown in Fig. \ref{Fig:06}, from which one can observe that the order of the three methods based on kappa index is different compared with that based on NMI. We also conduct more analysis to better understand why the results of NMI and kappa index are different.

\begin{figure}[!ht]
\centering
\includegraphics[height=50mm,width=80mm]{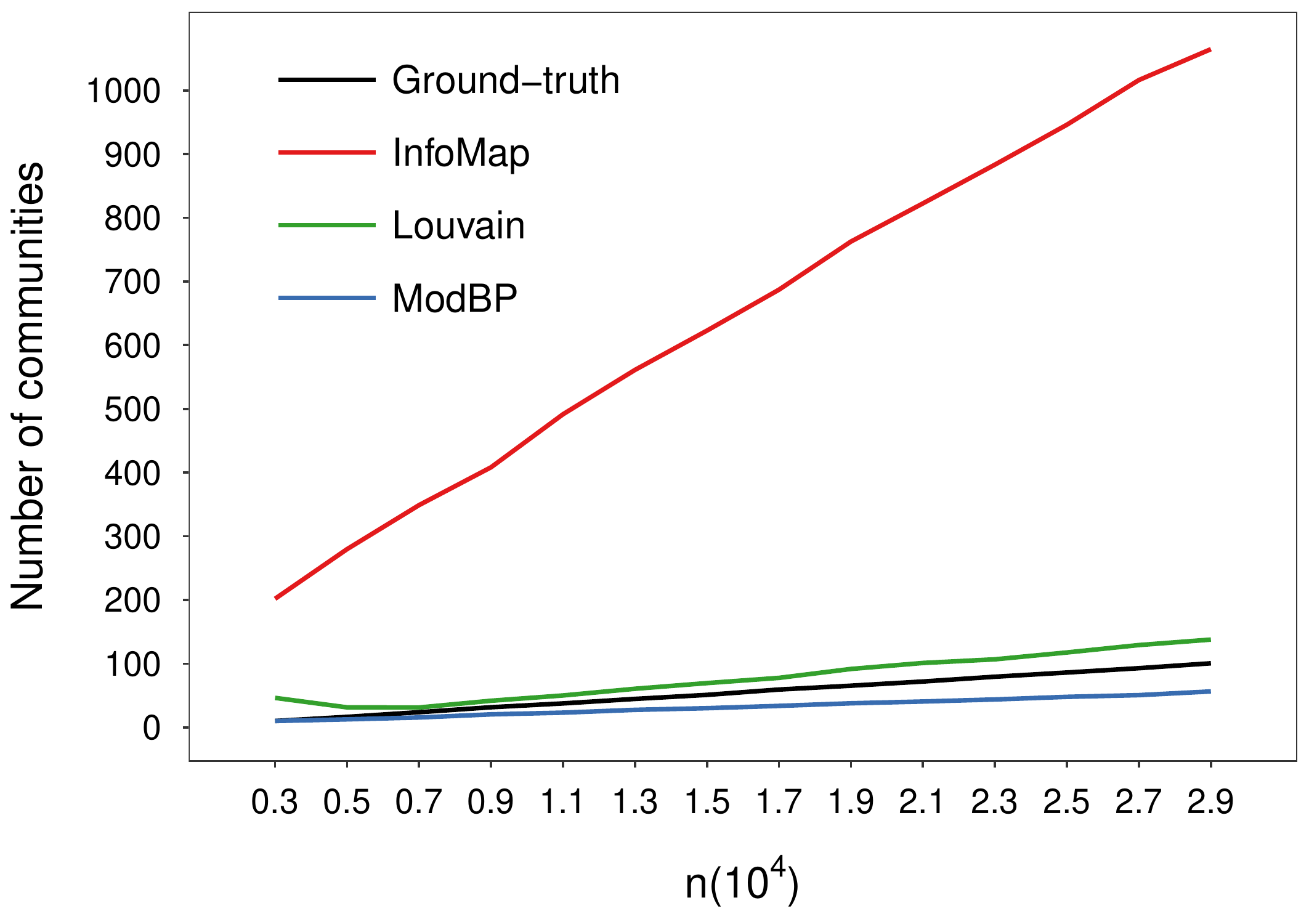}
\caption{Numbers of communities detected by three methods, InfoMap, Louvain and Modularity BP on LFR benchmarks with different network sizes. The parameters of networks are: average degree $k = 8$, mixing parameter $\mu = 0.45$, exponent of degree distribution $\gamma = 2$, exponent of community size distribution $\beta = 1$, maximum degree is $50$, and community sizes range
from $200$ to $400$.}\label{Fig:09}
\end{figure}

\begin{figure*}[ht!]
\centering
\includegraphics[height=80mm,width=140mm]{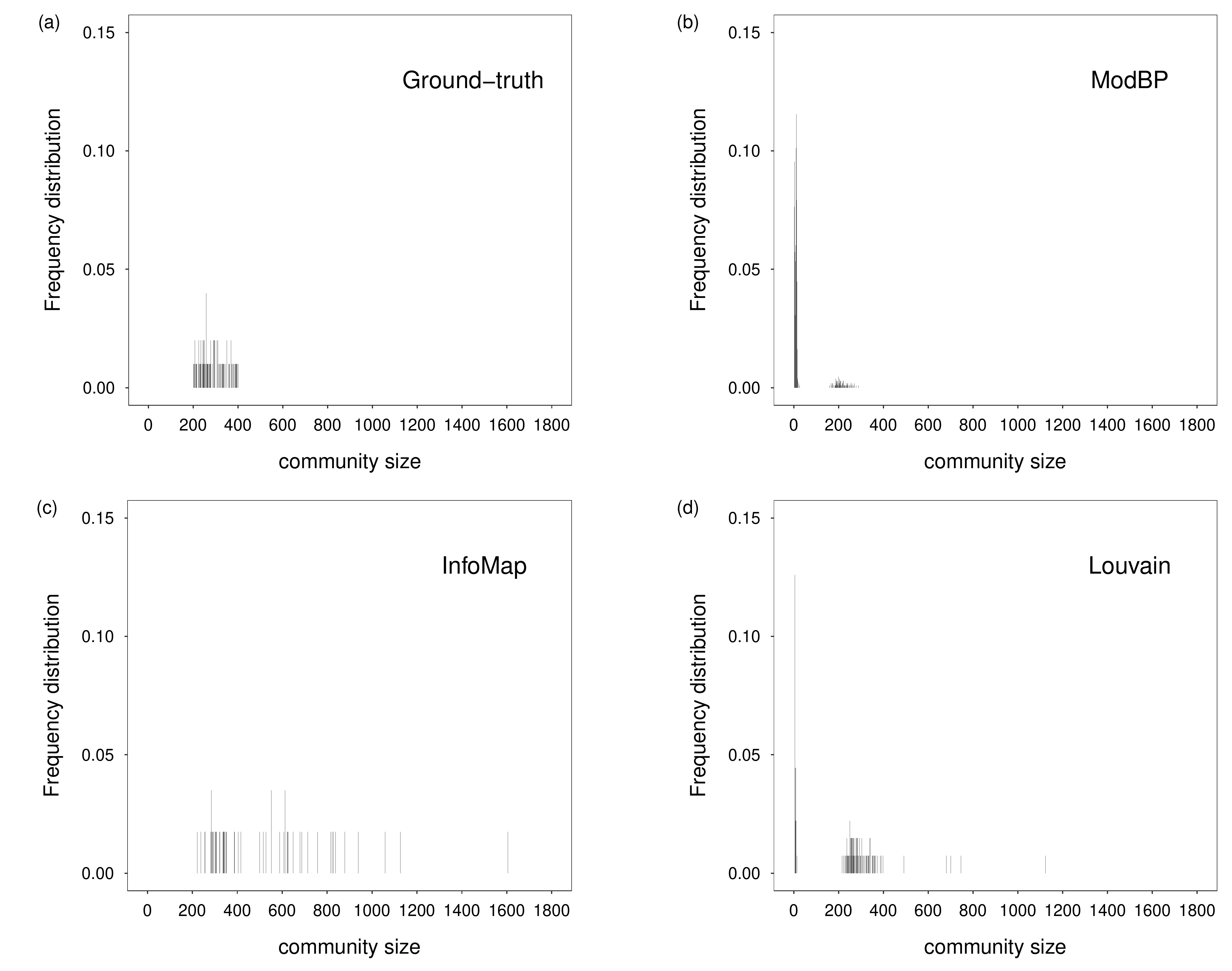}
\caption{Distributions of community sizes detected by three methods, InfoMap, Louvain and Modularity BP on LFR benchmark with network size $n = 29000$. The parameters of the network are: average degree $k = 8$, mixing parameter $\mu = 0.45$, exponent of degree distribution $\gamma = 2$, exponent of community size distribution $\beta = 1$, maximum degree is $50$, and community sizes range
from $200$ to $400$.}\label{Fig:10}
\end{figure*}

Fig. \ref{Fig:09} gives the numbers of communities detected by the three methods. Note that for ModBP, the community numbers are predefined as ground-truth, and the results of ModBP in the figure are numbers of nonempty communities. To take a closer look at the results, we use one network with size $n=29000$ as a case study, and the distributions of community sizes are shown in Fig. \ref{Fig:10}.

The conclusions from Fig. \ref{Fig:09} and Fig. \ref{Fig:10} are as follows: 1) The numbers detected by InfoMap are far more than those of the ground-truth, leading to higher NMI due to the finite size effect. 2)  Although the numbers of nonempty communities detected by ModBP are very close to those of the ground-truth, the distribution of community sizes is obviously different from that of the ground-truth. 3) The distribution of Louvain is the nearest one to that of the ground-truth. 4) The order of the three methods based on kappa index is more reasonable, indicating that kappa index is necessary for evaluation of community detection methods.

\begin{figure*}[ht]
\centering
\includegraphics[height=50mm,width=140mm]{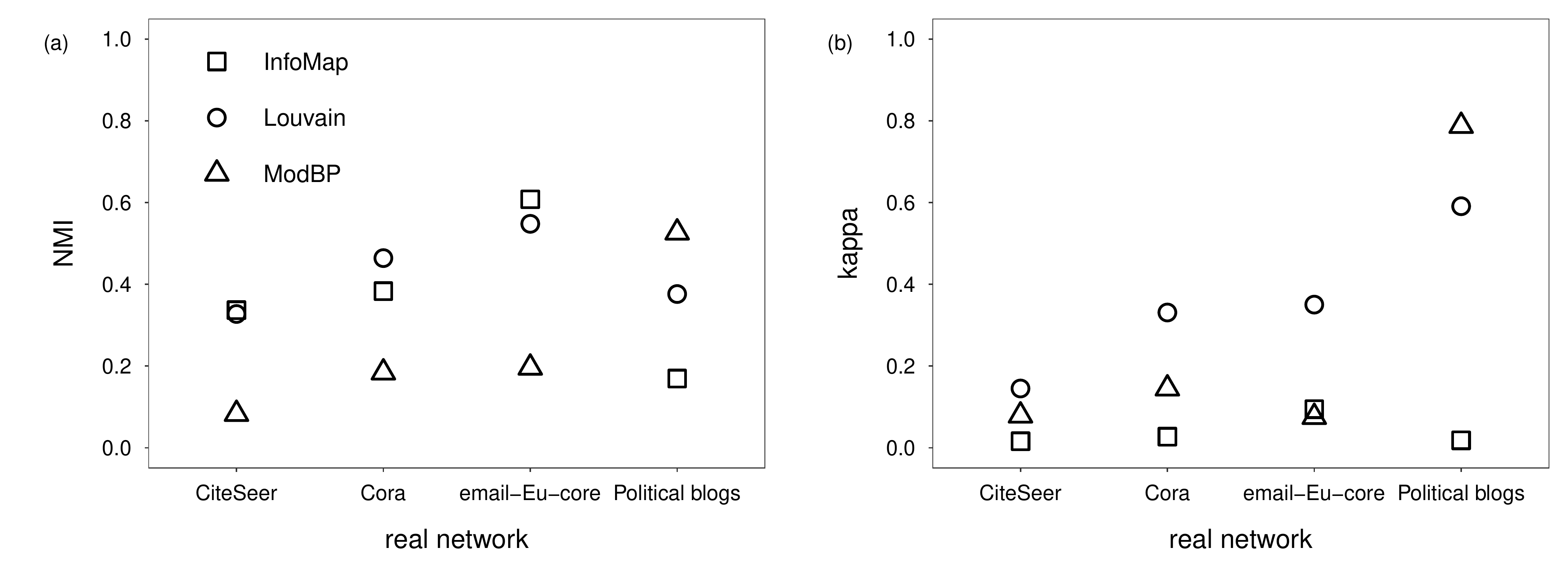}
\caption{Normalized Mutual Information and kappa index for three methods, InfoMap, Louvain and Modularity BP on real networks. The order of the three methods based on kappa is different compared with that based on NMI.}\label{Fig:08}
\end{figure*}

In addition, we have conducted more experiments on real networks, and the numerical results of NMI and kappa are given in Fig. \ref{Fig:08}, from which one can observe that: 1) The order of the three methods based on kappa is different compared with that based on NMI. 2) The order of InfoMap and ModBP is reversed, and overall, Louvain does well in the ranking. 3) The values of ModBP are better than InfoMap on real networks, and this is consistent with the results on synthetic networks with small network sizes $n$.


\section{Conclusion}\label{conclusion}
In this paper, we firstly review the metrics for evaluation of clustering methods, especially the ones for community structures detection. In summary, NMI-type metrics have systematic bias, ignore importance of small communities, and cannot evaluate the methods' ability to detect a single specific community of interest. To overcome these challenges, our proposed method maps the computed community labels to the ground-truth ones through integer linear programming, and then uses kappa index and F-score as evaluation metrics, instead of NMI. Systematic experimental results demonstrate the advantages of the method. There are several interesting problems for future work, including combining internal indices with external ones for performance evaluation in unsupervised learning and supervised learning, extending the method to evaluate general clustering results and to evaluate overlapping and hierarchical community structures, etc.

\ifCLASSOPTIONcompsoc
  \section*{Acknowledgments}
\else
  \section*{Acknowledgment}
\fi

This work was supported by Program for Innovation Research in Central University of
Finance and Economics.

\ifCLASSOPTIONcaptionsoff
  \newpage
\fi



\bibliographystyle{IEEEtran}
\bibliography{reference}
%



%
\end{document}